\documentclass[fleqn,usenatbib]{mnras}
\usepackage[T1]{fontenc}
\usepackage{ae,aecompl}
\usepackage{xtab}
\usepackage{courier}
\usepackage{multirow}
\usepackage{times,txfonts}
\usepackage{upgreek}
\usepackage{graphicx}
\usepackage{epstopdf}
\usepackage{amssymb}
\usepackage[section]{placeins}
\usepackage{caption}

\newcommand{\uHz}{$\mathrm{\mu}$Hz}

\title[Mode trapping in KIC\,10001893]{KIC\,10001893: A pulsating sdB star with multiple trapped modes}

\author[M.~Uzundag et al.]
{M.~Uzundag,$^{1}$\thanks{E-mail: Murat.Uzundag@ankara.edu.tr}
A.S. Baran,$^{2}$
R.H. {\O}stensen,$^{3}$
M.D. Reed,$^{3}$
J.H. Telting,$^{4}$\newauthor and
B.K. Quick$^{3}$
\\
$^{1}$Ankara \"{U}niversitesi, Fen Bilimleri Enstit\"{u}s\"{u}, Astronomi ve Uzay Bilimleri Anabilim Dal\i, \.{I}rfan Ba\c{s}tu\u{g} Caddesi 06110 D\i\c{s}kap\i, Ankara, Turkey\\
$^{2}$Uniwersytet Pedagogiczny, Obserwatorium na Suhorze, ul. Podchor\c{a}\.zych 2, 30-084 Krak\'ow, Polska\\
$^{3}$Department of Physics, Astronomy, and Materials Science, Missouri State University, Springfield, MO 65897, USA\\
$^{4}$Nordic Optical Telescope, Rambla Jos{\'e} Ana Fern{\'a}ndez P{\'e}rez 7, 38711 Bre{\~n}a Baja, Spain
}

\date{Accepted XXX. Received YYY; in original form ZZZ}
\pubyear{2017}

\begin{document}
\label{firstpage}
\pagerange{\pageref{firstpage}--\pageref{lastpage}}
\maketitle

\begin{abstract}
KIC\,10001893 is a V1093\,Her type pulsating subdwarf-B star, which was observed extensively by the {\it Kepler} spacecraft. It was a part of the survey searching for compact pulsators in the {\it Kepler} field of view. An amplitude spectrum of the star demonstrates a rich content of g-modes between 102 and 496\,$\mu$Hz as well as a few p-modes above 2000\,\uHz. In total, we found 110 frequencies. The g-mode region contains 104 frequencies, while the p-mode region contains just six, altogether indicating the hybrid nature of KIC\,10001893. The main goal of our analysis was to identify the detected modes and to find some features, which will significantly help modeling of the star. We found no multiplets, which points at a pole-on orientation, however, we defined modal degrees and relative radial orders using asymptotic period spacing. Eventually, we assigned 32 dipole {\it l}\,=\,1 and 18 quadrupole {\it l}\,=\,2 modes. The most remarkable feature we found are trapped modes, which are clearly seen in a reduce period diagram. It is the first time that three trapped modes are detected in one pulsating sdB star. Since the more trapped modes we find, the better sounding of the stellar interior we can do, this feature provides important constraints on the physical structure of the star. Mode trapping is likely caused by the He-H transition region and therefore it provides crucial constraints for making realistic theoretical models of hot subdwarfs.
\end{abstract}


\section{Introduction}
Subdwarf B (sdB) stars are located between the horizontal branch (HB) and the white dwarf cooling track, on the so-called 
extreme horizontal branch \citep[EHB,][]{heber2016}. The sdB stars are compact objects with masses typically around 0.5 solar masses and radii between 0.15 and 0.35 solar radii. Observed effective temperatures range from about 20,000 to 40,000\,K. These stars are core He-burning with thin hydrogen envelopes ($M_{env}$\,<\,0.01$M_{\odot}$). Such a small mass of the hydrogen envelope does not allow sdB stars to ascend the asymptotic giant branch. Therefore, after all helium is exhausted in their cores, sdB stars move directly to the white dwarf cooling track. The most enigmatic part of this evolution is the envelope-stripping phase, during which almost the entire hydrogen envelope is removed. This must occur close to the tip of the RGB, just before the occurrence of the helium flash, after which the envelope contracts while the core expands, and before the star settles down in its stable helium-burning phase. The cause of the envelope stripping is most likely mass transfer involving a close companion to an sdB star \citep{han2002}, or mass loss from a significantly enhanced stellar wind \citep{yong2000}.

The first pulsating sdB was discovered by a South African group of astronomers \citep{Kilkenny1997} and enables us to probe the interiors of sdB stars using asteroseismology. It is a powerful tool, which allows us to sound stellar structure using natural vibrations. Pulsating sdB (sdBV) stars are classified as V361\,Hya or V1093\,Her stars based on their periods. The former class is dominated by p-mode frequencies higher than 2000\,$\mu$Hz while the latter class is dominated by g-modes, which are usually below 1000\,\uHz. Some sdBV stars show both kinds of modes and are therefore called hybrid stars \citep{baran2005,schuh2006}. The first theoretical models of sdBV stars was a paper by \cite{charpinet1996} followed by \cite[e.g.][]{charpinet1997,charpinet2000,fontaine2003}. A successful application of those models can be found in a number of papers \citep[e.g.][]{koen1999,charpinet2011,vangrootel2013}.

During the last six years significant headway has been made in the field of sdBV stars. The {\it Kepler} spacecraft delivered unprecedented data revealing features that were rarely or never seen in ground-based data. These features include rotational multiplets \citep[e.g.][]{baran2012a}, asymptotic period spacing \citep[e.g.][]{reed2011}, Doppler beaming \citep[e.g.][]{telting2012}, and mode trapping \citep{ostensen2014,foster2015,kern2017}.

The {\it Kepler} observations made it possible to resolve the g-mode region in the V1093\,Her stars, something that, so far, 
has not been possible with ground-based data. This had profound impact on the theoretical models, where physics related to mixing (especially diffusion and overshooting) are poorly constrained. The first results by \citet{reed2011} indicated that the asymptotic sequences were much smoother than predicted by the early theoretical works, in which stratification played a strong role and produced a sharp boundary in the H/He transition zone. Such sharp boundaries are very efficient in producing trapped modes, and it was speculated that the absence of trapped modes might indicate that stronger diffusion could reduce the impact of this boundary. However, \citet{charpinet2014} demonstrated that while the boundary indeed produces significant trapping at low radial order, modes of higher order (which are the ones that appear with high amplitude in the observations) become less sensitive to this transition layer as their wavelengths become long compared to the size of the boundary. But this discussion was abruptly turned on its head with the first discovery of trapped modes in KIC\,10553698A by \citet{ostensen2014}. With such a clear signature of mode trapping it became possible for the modelers to test the adequacy of their models in reproducing the observed features. The first result was reported already by \citet{constantino2015}, where they explored models with different degrees of core overshooting parameters and found that their models could reproduce similar trapping effects caused by the sharp composition gradient at the edge of the partially mixed zone associated with the C-O/He boundary outside the convective inner core. More recently, \citet{ghasemi2017} explored models with various combinations of overshooting and diffusion parameters with the particular aim of matching the trapping patterns seein in KIC\,10553698A. In their scenarios with small and moderate overshooting the mixing leads to the emergence of convective shells around the core which produces mode trapping patterns that have comparable trapping signatures to those observed in KIC\,10553698A.

In this paper we present our analysis of KIC\,10001893, which was observed with the {\it Kepler} spacecraft. It was confirmed as a V1093\,Her pulsator with one month of data (Q3.3) during the {\it Kepler} survey phase \citep{ostensen2010,ostensen2011,baran2011}. \cite{baran2011} found 27 frequencies, mostly in the g-mode region, with asymptotic sequences as reported by \citet{reed2011}. \citet{silvotti2014} used data covering Q6--Q17.2 (1051\,days), which is the entire (continuous) data coverage available. However, they concentrated only on the lowest frequency region, below the so-called {\it cut-off} frequency and interpreted three frequencies, found in that region, as indications of exoplanets.

Here we have used the same 3-year dataset as \cite{silvotti2014}, and we provide a detailed analysis of the frequencies above the cut-off frequency. The increased span of the observations gives better resolution and lowers the detection limit compared to \cite{baran2011}, allowing us to detect more frequencies. We apply seismic tools to the pulsations to identify features useful for constraining models, in particular the \'echelle and reduced period diagrams. Our mode identifications seem to be reasonable, even though multiplets are absent, and the detection of trapped modes is particularly noteworthy.

\section{Photometric Data}
KIC\,10001893 was observed by the {\it Kepler} spacecraft. The observations started on 24 June 2010 and finished on 11 May 2013, which covered Q6 to Q17.2. We downloaded all available data from the ``Barbara A. Mikulski Archive for Space Telescopes'' (MAST)\footnote{archive.stsci.edu}. The {\it Kepler} spacecraft has two types of exposure times, which are short-cadence (SC, $\sim$1\,min), and long-cadence (LC, $\sim$30\,min). We used short-cadence (SC) data which covers the frequency range up to the Nyquist at 8495\,\uHz, and assuring both {\it p}- and {\it g}-mode regions are covered. The MAST database provides data processed in a couple of different ways (RAW, SAP, PDCSAP). We used fluxes with the pre-search data conditioning (PDCSAP) module and widely known as PDC fluxes. These fluxes are corrected for systematic errors and contamination from nearby stars, while the other fluxes are not. We clipped data at 4$\sigma$ and de-trended each monthly chunk of data separately. Finally, we stitched the monthly chunks together making the data ready for Fourier analysis.

\section{Fourier Analysis}
We used a Fourier technique to calculate the amplitude spectrum from Q6--17.2 data and we show it in Fig.\,\ref{fig_ft}. The frequency resolution equals 0.0162\,\uHz as defined by 1.5/T, where T is the time coverage of the data. We first tried the pre-whitening technique to fit and remove frequencies, however, profiles of signals at many frequencies were complex and we were unable to fit those. Therefore, we decided to determine the frequencies by-eye. We applied a 5-$\sigma$ detection threshold \citep{baran2015a} of 0.02\,ppt. We checked the detected frequencies against the list of artifacts \citep{baran2013}. Eventually, we found 110 significant peaks above the detection threshold, showing KIC\,10001893 to be a rich g-mode pulsator with most pulsations below 500\,$\mu$Hz but significant frequencies all the way up to 3897\,\uHz. We list all detected frequencies in Table\,\ref{tab1}.

\subsection{p-modes}
\begin{figure}
\centering
\includegraphics[scale=0.45]{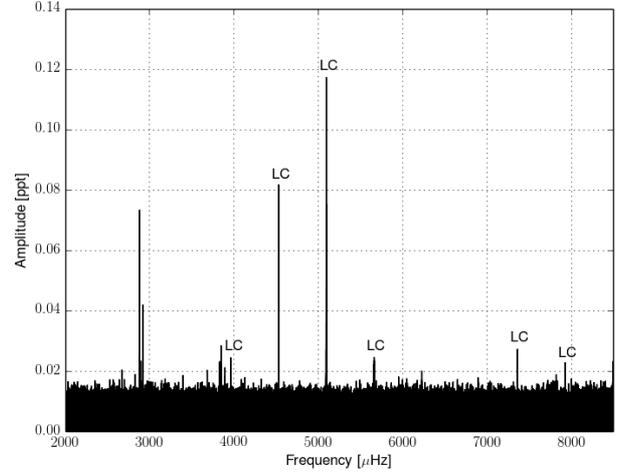}
\caption{A close-up of the amplitude spectrum showing the high-frequency region.}
\label{fig_p}
\end{figure}

We found six frequencies in the high-frequency region. This area is relatively sparse in frequencies, with all of them located between 2880 and 3900\,\uHz. We show a close-up of this region in Fig.\,\ref{fig_p}. The highest amplitude is 0.07\,ppt.
We assigned those frequencies with p-modes. No multiplets are present in this region, and since such features are one of the most useful tools to make mode identifications, we were left with no chance to directly identify individual modes. The multiplets can also be useful in estimating stellar rotation. An absence of split modes can be caused by low S/N in this region or if the stellar rotation axis is pointed towards the observer, which efficiently suppresses the amplitudes of the side components, or if the rotation rate is so slow that the span of the data is insufficient to resolve the multiplets. More details will be given in Section\,\ref{multiplets}. A few frequencies beyond 4000\,$\mu$Hz are of instrumental origin, related to the widely known LC-readout time.

\subsection{g-modes}
\begin{figure}
 \centering
 \includegraphics[scale=0.45]{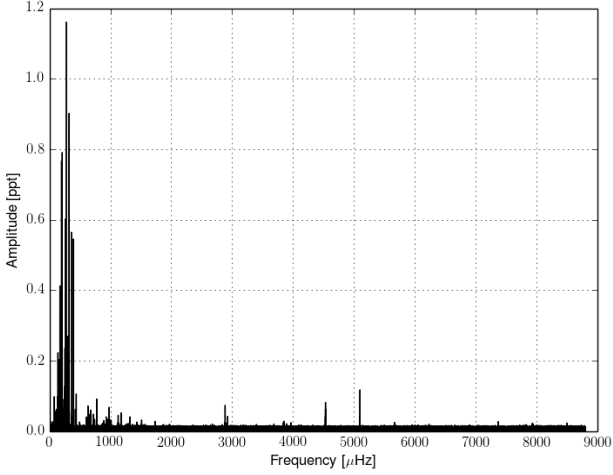}
 \caption{Amplitude spectrum, calculated up to the Nyquist frequency, showing both g- and p-modes. Close-ups of both regions are shown in Figs.\,\ref{fig_p} and \ref{fig_g}.}
 \label{fig_ft}
\end{figure}

\begin{figure}
\centering
\includegraphics[scale=0.45]{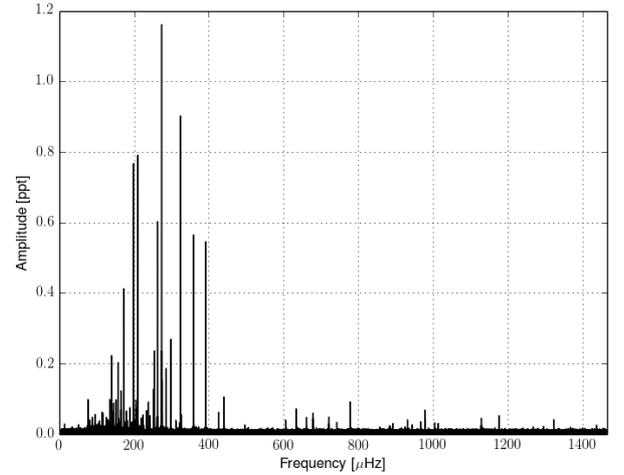}
\caption{A close-up of the amplitude spectrum showing the low-frequency region}
\label{fig_g}
\end{figure}

We show the g-mode region of the amplitude spectrum in Fig.\,\ref{fig_g}. This region is rich in frequencies with relatively large amplitudes compared to the p-modes (see Fig.\,\ref{fig_ft}). The frequencies with the highest amplitudes are located below 400\,\uHz. The region above 500\,$\mu$Hz up to 2000\,$\mu$Hz shows numerous low amplitude peaks which do not follow the spacing sequences of the low order modes. Although modes are not expected to follow the asymptotic spacing at low $n$, there seems to be too many of them for all to be of degree {\it l}\,=\,1 or 2. 
In several other Kepler-observed sdBV stars \cite{telting2014,foster2015,kern2017},  multiplets in this frequency range indicate 3\,$\le$\,{\it l}\,$\le$\,9 modes. A few frequencies are of instrumental nature (566.43\,$\mu$Hz and its multiples). In total, we found 24 frequencies in this region. But since we cannot see any signs of rotational splitting, there is no chance of identifying these modes.

\vspace{10pt}
\tablecaption{A list of frequencies found in the amplitude spectrum of KIC\,10001893. Trapped modes are indicated with a ``t'' letter.}
\begin{xtabular}{c c c c c c}
	\hline \hline
	ID & Freq[\uHz] & Period[s] & Ampl[ppt] & $l$ & $n$ \\
	\hline
	$f_{1}$ &76.45610 &13079.4 &0.026 & & \\
	$f_{2}$ &77.52959 &12898.3 &0.098 & & \\
	$f_{3}$ &78.45598 &12746.0 &0.029 & & \\
	$f_{4}$ &79.71621 &12544.5 &0.023 & & \\
	$f_{5}$ &81.97998 &12198.1 &0.040 & & \\
	$f_{6}$ &83.93838 &11913.5 &0.021 & & \\
	$f_{7}$ &87.36142 &11446.7 &0.024 & & \\
	$f_{8}$ &89.08844 &11224.8 &0.048 & & \\
	$f_{9}$ &91.74985 &10899.2 &0.024 & & \\
	$f_{10}$ &96.93867 &10315.8 &0.056 & & \\
	$f_{11}$ &98.98050 &10103.0 &0.020 & & \\
	$f_{12}$ &102.17218 &9787.4 &0.027 &1 &36 \\
	$f_{13}$ &105.01995 &9522.0 &0.031 &1 &35 \\
	$f_{14}$ &107.83350 &9269.3 &0.031 & & \\
	$f_{15}$ &108.18998 &9243.0 &0.036 &1 &34 \\
	$f_{16}$ &111.64452 &8957.0 &0.023 & & \\
	$f_{17}$ &114.41647 &8740.0 &0.062 & & \\
	$f_{18}$ &114.71047 &8717.6 &0.038 &1 &32 \\
	$f_{19}$ &116.91257 &8553.4 &0.060 & & \\
	$f_{20}$ &118.33339 &8450.7 &0.032 &1 &31 \\
	$f_{21}$ &120.49499 &8299.1 &0.022 & & \\
	$f_{22}$ &122.07627 &8191.6 &0.019 &1 &30 \\
	$f_{23}$ &123.88963 &8071.7 &0.023 & & \\
	$f_{24}$ &125.85581 &7945.6 &0.048 &1 &29 \\
	$f_{25}$ &126.89067 &7880.8 &0.034 &1 &t \\
	$f_{26}$ &130.55000 &7659.9 &0.041 &1 &28 \\
	$f_{27}$ &132.47489 &7548.6 &0.033 & & \\
	$f_{28}$ &135.32898 &7389.4 &0.098 &1 &27 \\
	$f_{29}$ &140.52048 &7116.4 &0.223 &1 &26  \\
	$f_{30}$ &141.47473 &7068.4 &0.029 & & \\
	$f_{31}$ &145.73647 &6861.7 &0.036 & & \\
	$f_{32}$ &146.10057 &6844.6 &0.088 &1 &25 \\
	$f_{33}$ &152.06118 &6576.3 &0.098 &1 &24 \\
	$f_{34}$ &155.96730 &6411.6 &0.028 & & \\
	$f_{35}$ &158.08277 &6325.8 &0.204 &1 &23 \\
	$f_{36}$ &160.05890 &6247.7 &0.048 &1 &t \\
	$f_{37}$ &163.05499 &6132.9 &0.033 & & \\
	$f_{38}$ &164.99199 &6060.9 &0.069 & & \\
	$f_{39}$ &165.39587 &6046.1 &0.123 &1 &22 \\
	$f_{40}$ &166.21235 &6016.4 &0.045 & & \\
	$f_{41}$ &172.45839 &5798.5 &0.063 & & \\
	$f_{42}$ &172.72052 &5789.7 &0.413 &1 &21 \\
	$f_{43}$ &173.19016 &5774.0 &0.057 & & \\
	$f_{44}$ &176.67844 &5660.0 &0.034 & & \\
	$f_{45}$ &178.79492 &5593.0 &0.021 & & \\
	$f_{46}$ &180.55430 &5538.5 &0.065 &1 &20 \\
	$f_{47}$ &181.94388 &5496.2 &0.039 &2 &35 \\
	$f_{48}$ &187.26591 &5340.0 &0.019 &2 &34 \\
	$f_{49}$ &189.09310 &5288.4 &0.076 &1 &19 \\
	$f_{50}$ &192.71535 &5189.0 &0.033 &2 &33 \\
	$f_{51}$ &196.41734 &5091.2 &0.036 & & \\
	$f_{52}$ &199.11196 &5022.3 &0.770 &1 &18 \\
	$f_{53}$ &201.17486 &4970.8 &0.034 &1 &t \\
	$f_{54}$ &204.69571 &4885.3 &0.097 &2 &31 \\
	$f_{55}$ &210.68155 &4746.5 &0.793 &1 &17 \\
	$f_{56}$ &211.18878 &4735.1 &0.040 &2 &30 \\
	$f_{57}$ &217.75580 &4592.3 &0.025 &2 &29 \\
	$f_{58}$ &219.40891 &4557.7 &0.046 &2 &t \\
	$f_{59}$ &224.71910 &4450.0 &0.054 &1 &16 \\
	$f_{60}$ &225.89681 &4426.8 &0.027 &2 &28 \\
	$f_{61}$ &234.17558 &4270.3 &0.066 &2 &27 \\
	$f_{62}$ &238.98288 &4184.4 &0.091 &1 &15 \\
	$f_{63}$ &243.14335 &4112.8 &0.052 &2 &26 \\
	$f_{64}$ &252.79336 &3955.8 &0.129 &2 &25 \\
	$f_{65}$ &255.16713 &3919.0 &0.236 &1 &14 \\
	$f_{66}$ &262.97796 &3802.6 &0.602 &2 &24 \\
	$f_{67}$ &273.33606 &3658.5 &0.086 &2 &23 \\
	$f_{68}$ &274.30326 &3645.6 &1.162 &1 &13 \\
	$f_{69}$ &276.35760 &3618.5 &0.053 &2 &t \\
	$f_{70}$ &286.00028 &3496.5 &0.186 &2 &22 \\
	$f_{71}$ &298.59659 &3349.0 &0.051 &2 &21 \\
	$f_{72}$ &298.66794 &3348.2 &0.270 &1 &12 \\
	$f_{73}$ &312.06116 &3204.5 &0.039 &2 &20 \\
	$f_{74}$ &323.98107 &3086.6 &0.902 &1 &11 \\
	$f_{75}$ &326.78670 &3060.1 &0.053 &2 &19 \\
	$f_{76}$ &359.68635 &2780.2 &0.565 &1 &10 \\
	$f_{77}$ &391.37411 &2555.1 &0.546 &1 &9 \\
	$f_{78}$ &426.03953 &2347.2 &0.062 & & \\
	$f_{79}$ &440.50922 &2270.1 &0.105 &1 &8 \\
	$f_{80}$ &496.15480 &2015.5 &0.026 &1 &7 \\
	$f_{81}$ &605.36352 &1651.9 &0.040 & & \\
	$f_{82}$ &633.75372 &1577.9 &0.072 & & \\
	$f_{83}$ &660.89485 &1513.1 &0.048 & & \\
	$f_{84}$ &679.02492 &1472.7 &0.060 & & \\
	$f_{85}$ &721.03251 &1386.9 &0.049 & & \\
	$f_{86}$ &742.05995 &1347.6 &0.034 & & \\
	$f_{87}$ &777.84691 &1285.6 &0.091 & & \\
	$f_{88}$ &883.93883 &1131.3 &0.022 & & \\
	$f_{89}$ &892.29945 &1120.7 &0.030 & & \\
	$f_{90}$ &924.72720 &1081.4 &0.021 & & \\
	$f_{91}$ &930.92533 &1074.2 &0.041 & & \\
	$f_{92}$ &943.66330 &1059.7 &0.026 & & \\
	$f_{93}$ &966.09023 &1035.1 &0.035 & & \\
	$f_{94}$ &977.51710 &1023.0 &0.068 & & \\
	$f_{95}$ &1004.52034 &995.5 &0.032 & & \\
	$f_{96}$ &1013.17122 &987.0 &0.030 & & \\
	$f_{97}$ &1128.41345 &886.2 &0.045 & & \\
	$f_{98}$ &1175.77895 &850.5 &0.052 & & \\
	$f_{99}$ &1267.58778 &788.9 &0.021 & & \\
	$f_{100}$ &1294.66597 &772.4 &0.022 & & \\
	$f_{101}$ &1321.87706 &756.5 &0.041 & & \\
	$f_{102}$ &1435.75017 &696.5 &0.026 & & \\
	$f_{103}$ &1513.31719 &660.8 &0.032 & & \\
	$f_{104}$ &1734.30454 &576.6 &0.028 & & \\
	$f_{105}$ &2884.33804 &346.7 &0.073 & & \\
	$f_{106}$ &2898.55072 &345.0 &0.023 & & \\
	$f_{107}$ &2925.68753 &341.8 &0.042 & & \\
	$f_{108}$ &3834.35582 &260.8 &0.023 & & \\
	$f_{109}$ &3849.11470 &259.8 &0.028 & & \\
	$f_{110}$ &3897.11613 &256.6 &0.021 & & \\
	\hline \hline
	\label{tab1}
\end{xtabular}

\section{Multiplets}
\label{multiplets}
The ultimate goal of our analysis was to identify modes to constrain theoretical models of this star. Mode identifications describe a mode's geometry, i.e. a radial order {\it n}, a modal degree {\it l} and an azimuthal order {\it m} of the mode. Information on these parameters is very crucial for calculation of stellar interior. Each mode is described by three numbers and therefore each mode adds three unknown parameters to the model. Since the more free parameters there are, the less reliable the model is, it is important to identify as many modes as possible.

In the presence of stellar rotation, non-radial modes of degree {\it l} split into 2{\it l}+1 components differing in {\it m} number. The azimuthal frequencies can be derived from the following equation

\begin{equation}
\nu_{n,l,m} = \nu_{n,l,0} + \Delta \nu_{n,l,m} = \nu_{n,l,0} + m \frac{1- C_{n,l}}{P_{\rm rot}}
\end{equation}

where $\Delta \nu_{n,l,m}$ is a rotational splitting. P$_{\rm rot}$ is a star's rotation period and $C_{n,l}$ is the Ledoux constant. We searched for multiplets among g-modes with a null result. The most likely explanation for this is that the inclination is too low, as was inferred by \cite{silvotti2014}. The authors concluded that the inclination may be as low as just a few degrees. If so, KIC\,10001893 is another example of a pole-on oriented sdB star \citep[such as KIC\,8302197][]{baran2015b}. Other possibilities, like extremely slow rotation or selective driving of m\,=\,0 modes cannot be completely ruled out, though we consider this to be unlikely. The lack of multiplets prevents us from identifying azimuthal orders and determining a rotation rate for KIC\,10001893.

If the absence of rotational splitting is due to extremely slow rotation, the time resolution of the Kepler data set provides a limit on the rotation period of Prot\,$\ge$\,715\,days. Here we used the FWHM of a few amplitude-stable modes (0.00116\,c/d) to approximate the resolution needed to tell apart an {\it l}>1 multiplet, and we assumed the $\delta$m splitting of (1-C$_{\rm nl}$)/P$_{\rm rot}\ge$0.83/P$_{\rm rot}$, expected for high-order {\it l}>1 g-modes. We examined the amplitudes of many of our modes and do not see a single characteristic time scale of amplitude variability longer than 715\,days, which one may expect for unresolved beating between multiplet components. Therefore, if the lack of multiplets is due to very slow rotation, the rotation period must be well longer than the length of the dataset. This is in contrast with typical rotational rates of tens of days detected thus far \citep[e.g.][]{baran2012a,baran2012b,telting2012,ostensen2014,reed2014,foster2015}.

For the other known case of an sdB pulsator which does not show any multiplets in a 3-year dataset, KIC\,8302197, \citet{baran2015b} argued that the lack of multiplets can be due to geometrical cancellation if the inclination is lower than a few degrees, but that the probability of having such low inclination angles is such that one would expect only a few such cases in every 1000 sdB stars. Finding two such cases in a sample of only $\sim$16 well-studied sdB pulsators, is therefore somewhat problematic.

\section{Period Spacing}
In the asymptotic limit {\it n}\,\texttt{>{}>}\,{\it l}, consecutive radial overtones are evenly spaced in period \citep[e.g.][]{reed2011}. For given {\it n} and {\it l} values, periods of consecutive overtones can be calculated from
\begin{equation}\label{eq:asym}
P_{l,n} = \frac{\Delta\Pi}{\sqrt{l(l+1)}} n + \epsilon_{l}
\end{equation}
where $\Delta\Pi$ is the reduced period spacing and $\epsilon_{l}$ is a constant offset smaller than $\Delta\Pi$. With this equation we define the asymptotic radial order of a mode, $n$, but note that this does not allow us to derive the true radial order of a mode, which is usually denoted $k$, since any trapped modes will be inserted into the the mode sequence. Thus, each trapped mode will bump the mode order $k$ by one relative to the asymptotic order $n$, at the point in the sequence where it occurs. The reduced period spacing $\Delta\Pi$ can be inferred from the \'echelle diagrams by converting the observed period spacings for each {\it l}, $\Delta P_{l}$, to reduced period, $\Delta\Pi$\,=\,$\sqrt{l(l+1)}\Delta P_{l}$. The offset $\epsilon_l$ can also be derived from the \'echelle diagrams as the mean value of the observed periods associated with a given {\it l} modulo the period spacing.
It was found from observations that the period spacing of {\it l}\,=\,1 modes is around 250\,s \citep{reed2011}, matching  predictions from theoretical models \citep{charpinet2000}. The spacings of {\it l}\,=\,2 modes are found to be around 145\,s, so that both produces reduced spacings of $\sim$350\,s. Period spacings for {\it l}\,=\,3 would translate to around 102\,s, but has never been detected.

\begin{figure*}
\centering
\includegraphics[width=\hsize]{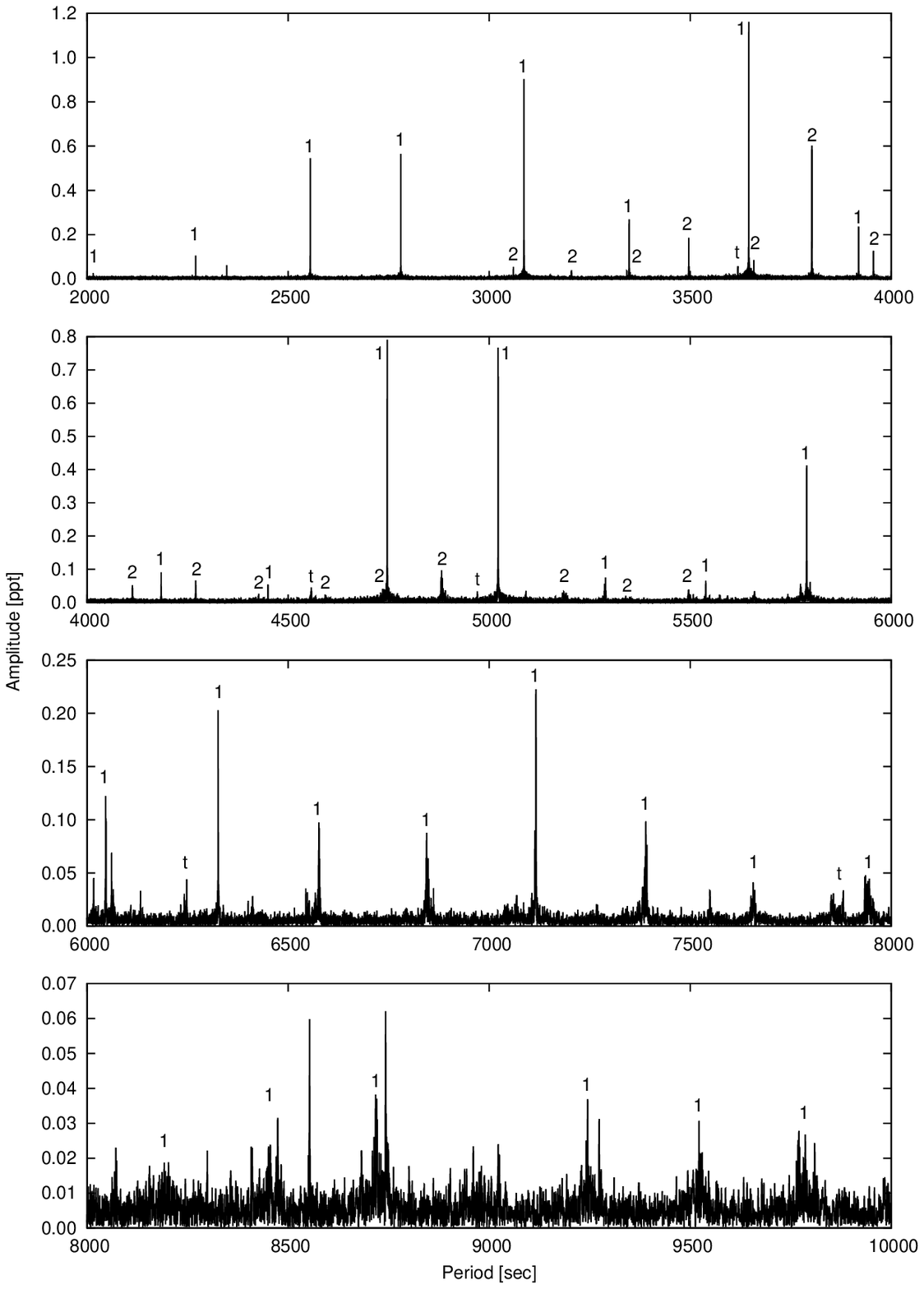}
\caption{Modal degree assignment. Note that the horizontal axis is in period instead of frequency.}
\label{fig_ps}
\end{figure*}

Since pulsation periods depend on the local sound speed, chemical stratification in the resonant cavity will cause fluctuations in the observed period spacings at low $n$, but these should disappear at higher $n$ \citep{charpinet2014}. For KIC\,10001893 we derived the average of the spacing value for {\it l}\,=\,1 to be 268.0$\pm$0.5\,s, and 153$\pm$0.4\,s, respectively for {\it l}\,=\,1 and 2. Our {\it l}\,=\,1 and 2 values match those of \citet{reed2011}. The sequence of {\it l}\,=\,1 modes is fairly complete, with 32 overtones identified. While 18 {\it l}\,=\,2 overtones were identified, the sequence is nearly complete, only missing {\it n}\,=\,32. The identification of modal degrees is marked in Fig.\,\ref{fig_ps} and provided in Table\,\ref{tab1}. KIC\,10001893 is therefore another very successful case for mode identification in sdB stars, providing significant constraints for stellar modeling. Although the multiplets are not present, their absence actually makes it simpler since we can associate all modes with {\it m}\,=\,0 and avoid the ambiguity which exists when having incomplete multiplets, as was seen in KIC\,10553698A \citep{ostensen2014}. A pole on orientation means that only the central components are visible, while an extremely slow rotation would imply that modes of different $m$ have the same frequency, effectively producing the same result. In either case, this does not reduce the value of the identification that we are able to make from asymptotic sequences alone. Although, we would have liked to use the multiplets for confirmation of our mode designations, having obtained those verifications from a large number of cases already, we are confident that the asymptotic sequences alone are sufficient to derive reliable mode identifications.

\cite{reed2011} found period spacings between 230 and 270\,s for the g-mode pulsators in the {\em Kepler} field. The spacing is determined by the size of the pulsation resonant cavity and is the first asteroseismic diagnostic that can be used to constrain pulsation models. For instance \cite{schindler2015} found that models computed with MESA produced much larger cores when type II opacities were included and either overshoot or diffusion were turned on. Still, such large cores would reduce the size of the propagation cavity for g-modes and these are incompatible with all but the shortest period spacings found by \cite{reed2011}. \citet{constantino2015} also concluded that their models tended to produce period spacings significantly below the average inferred from observations. The period spacing is therefore an important constraint for theoretical models, and since KIC\,10001893 has one of the largest period spacings for known g-mode pulsators, it is an important case for modeling the extent of resonant cavities.

\section{Mode Trapping}
The asymptotic relation for g-modes, Eq.~\ref{eq:asym}, holds for idealised stellar models with homogeneous composition. In that case the boundaries for a g-mode being a standing wave is a surface and, usually, a convective core. The consecutive overtones are perfectly spaced making an even sequence of modes of a given modal degree {\it l}, and the \'echelle diagrams would show a vertical ridge. However, real sdB stars are compositionally stratified. Composition discontinuities will produce steep gradients in density, which may act as a boundary for stellar pulsations. In the sdB stars, the two important composition transition regions are the H/He boundary and the deep transition region between the helium mantle and the central convective core. In the models of \cite{charpinet2014} it is the H/He transition that produces the strongest trapping effect, while in the work of \cite{ghasemi2017} strong trapping effects are produced by convective shells forming in the transition zone between the C/O-enhanced convective core and the radiative He mantle. These boundaries can cause partial reflection of the waves, or even act as an extra reflection surface which cause extra modes to be inserted into the otherwise roughly evenly spaced sequence of frequencies.

While the detection of trapped modes is in itself interesting, finding several of them and thereby being able to estimate the spacing between consecutive trapped modes, is even more useful since the spacing between consecutive trapped modes can
be a particularly important diagnostic when comparing observations with theoretical models. When searching for trapped modes it is convenient to start with the identifications made in the \'echelle diagram for the {\it l}\,=\,1 and 2 sequences and plot them in a reduced-period diagram (where period, $P$, is converted to reduced period $\Pi$\,=\,$P\sqrt{\ell\left(\ell\,+\,1\right)}$). After conversion to reduced period, modes of the same order should fall roughly on top of each other. In the \'echelle diagram, trapped modes are shifted off the ridge and cannot therefore be assigned to a particular {\it l}. But with some trial and error it is possible to use the fact that the reduced-period differences should be the same for different {\it l} to find solutions where two observed modes correspond to the same order $n$ for different $l$ and therefore line up in the reduced period diagram.

\begin{figure}
\centering
\includegraphics[scale=0.75]{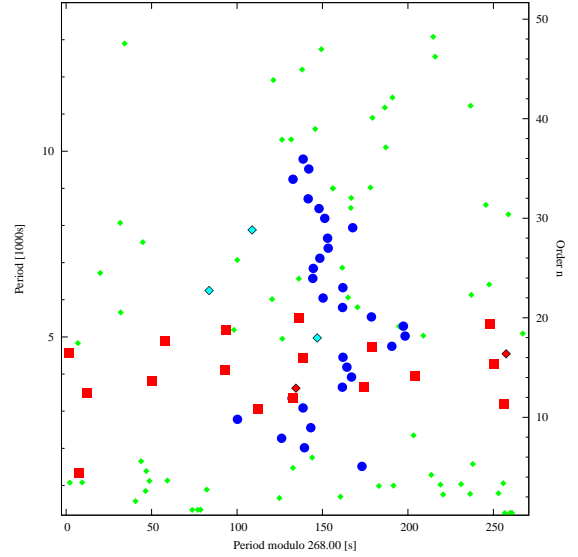}
\caption{\'Echelle diagram folded over the average {\it l}\,=\,1 period spacing. Blue dots represent dipole modes, red squares quadrupole modes, while diamonds with black outlines represent trapped modes. Green diamonds show unidentified modes.}
\label{fig_e1}
\end{figure}

\begin{figure}
\centering
\includegraphics[scale=0.75]{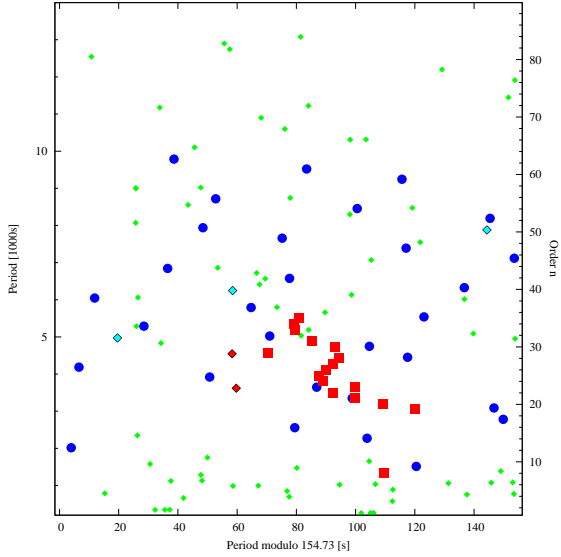}
\caption{Same as in Fig.\,\ref{fig_e1} but for quadrupole modes. }
\label{fig_e2}
\end{figure}

We calculated \'echelle diagrams for {\it l}\,=\,1 up to 5, though only {\it l}\,=\,1 and 2 overtones show evidence of systematic ridges (shown in Figs.\,\ref{fig_e1} and \ref{fig_e2}). The diagram of {\it l}\,=\,1 modes shows a ``hook'' feature observed in some other sdBV stars \citep[e.g.][]{baran2012b}. Such a feature appears in some models \citep[e.g. Fig.\,1 of][]{charpinet2014}, and is related to the chemical stratification, but a quantitative explanation of this feature has not been presented, therefore we will postpone a discussion of this feature until models for this star are available. We used the asymptotic relation to define {\it n} values for our modes, and these values are listed in Table\,\ref{tab1}. We note that there may be a zero-point offset {\it n$_l$} between the listed {\it n} values and the actual mode order {\it k}, an offset that we cannot determine without the appropriate modeling. The trapped modes are effectively inserted between consecutive asymptotic radial orders, implying smaller period spacings between adjacent radial orders around the trapped mode.

\begin{figure}
\centering
\includegraphics[scale=0.75]{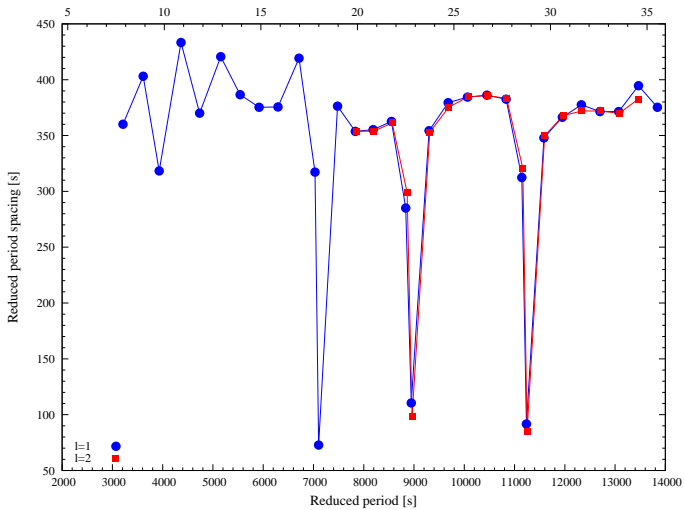}
\caption{Reduced Period Diagram. Reduced period is defined as $P\sqrt{\ell\left(\ell\,+\,1\right)}$.}
\label{fig_rpd}
\end{figure}

The identification of two trapped modes between radial orders {\it n}\,=\,22 and 23, and 28 and 29, is very convincing as they are well matched in both the {\it l}\,=\,1 and 2 sequences (Fig.\,\ref{fig_rpd}). The radial orders around the trapped modes agree between both modal degrees, as predicted by \cite{charpinet2000} (their Eq.~30). A possible third trapped mode around {\it n}\,=\,18 is only present in the {\it l}\,=\,1 sequence, and as we could not recover the period-spacing relation for {\it l}\,=\,2 in that 
region, the significance of that trapped mode is much less convincing. Indeed, there are other unidentified modes which could be associated with trapped modes, but without multiplets, we have no way of preferring one over the other. For the overlap region, however, the match is exceptionally good, which gives us high confidence that those modes have been correctly identified.

The reduced-period spacing between the trapped modes is roughly 2000\,s and this observation agrees with the third trapped mode. In the models of \cite{charpinet2000}, this spacing can be taken as a direct measure of the mass of the H envelope, and it was estimated by \cite{charpinet2000} that it should be of the order of 2000\,s, in excellent agreement with our observation. \cite{charpinet2000} found that for some trapped modes the minimum in period spacing does not correspond to the minimum of the kinetic energy, as expected, but to the maximum instead. Such trapped modes will not be spaced by around 2000\,s, but somewhat closer, and will not overlap between modal degrees. We found two spacings, both close to 2000\,s. According to \cite{charpinet2000}, our observations suggest that only the He/H transition zone contributes to the trapping, while the impact from the other, C-O/He, zone is negligible. Having the spacing measured precisely, we could compare it with the models to find the location of the base of the H envelope and its mass.This task will be done as soon as the specific model of this star is available.

\section{Summary and Conclusions}
We have analyzed KIC\,10001893, a V1093\,Her type sdBV star observed with the {\it Kepler} spacecraft. The data consists of a nearly uninterrupted 1051\,d sequence with 1-minute cadence. An amplitude spectrum uncovered a rich content of 110 pulsation modes. We recovered all 27 frequencies of \citep{baran2011}, and while amplitudes vary somewhat, those frequencies are consistent between Q3.3 and this much longer data set. The amplitude spectrum shows both g- and p-modes, but g-modes are strongly favored, with only 6 p-modes all below 0.1 ppt in amplitude. So while KIC\,10001893 is clearly a hybrid pulsator, it is quite different from the DW\,Lyn-type hybrid pulsators which show p- and g-modes in similar amounts. Hybridity can be an important feature which provides simultaneous constraints both on core and envelope parameters.

We find that KIC\,10001893 is quite an enigmatic object with respect to rotationally split multiplets, a feature normally easily seen in similar stars. We found no multiplets among either p- or g-modes, with the most likely explanation being a very low inclination of the star. This would then be the second sdBV star oriented pole on \citep{baran2015b}. We identified the modal degree for 50 of the detected g-modes, which is about half of the modes, and includes all of the modes with amplitudes higher than 0.1\,ppt. Out of those 50 modes, 32 are dipole modes and 18 are quadrupole modes. We calculated \'echelle diagrams for {\it l}\,=\,1 and 2 (Fig.\,\ref{fig_e1} and \,\ref{fig_e2}) and they show wavy excursions from a straight ridge, which are indicative of non-uniform density profiles in the stellar interior. This feature is similar to ones detected in other {\it Kepler}-observed sdBV stars \citep[see e.g.][]{baran2012b,ostensen2014,foster2015,baran2017}.

After identifying almost complete sequences of consecutive radial orders for both {\it l}\,=\,1 and 2, including several trapped modes, we were able to construct a reduced-period diagram showing almost perfect alignment of the two sequences. The location of the trapped modes, and particularly the spacing between them, provides a useful tool to examine the stellar interior and test various modeling prescriptions against each other. The trapping can be caused by either the transition region of He/H at the base of the H-rich envelope, or deeper trapping regions associated with the transition region between the convective and radiative part of the core. Fig.\,\ref{fig_rpd} shows three trapped modes, with two of them overlapping between {\it l}\,=\,1 and 2 sequences. This pattern is almost identical to the one seen in KIC\,10553698A, but with all three trapped modes shifted to slightly higher periods.
The overlap between the two sequences is so close that in spite of the lack of multiplets as a cross-check for mode identifications, we are confident of our conclusions, at least for the region of overlap. We consider KIC\,10001893 to be one of the most prominent sdB stars for applying asteroseismology to infer the physics of the deep interiors of core-helium burning stars.

\section*{Acknowledgements}
ASB gratefully acknowledges financial support from the Polish National Science Center under project No.\,UMO-2011/03/D/ST9/01914.

\bibliographystyle{mnras}
\bibliography{myrefs}

\begin{thebibliography}{}
\makeatletter
\relax
\def\mn@urlcharsother{\let\do\@makeother \do\$\do\&\do\#\do\^\do\_\do\%\do\~}
\def\mn@doi{\begingroup\mn@urlcharsother \@ifnextchar [ {\mn@doi@}
  {\mn@doi@[]}}
\def\mn@doi@[#1]#2{\def\@tempa{#1}\ifx\@tempa\@empty \href
  {http://dx.doi.org/#2} {doi:#2}\else \href {http://dx.doi.org/#2} {#1}\fi
  \endgroup}
\def\mn@eprint#1#2{\mn@eprint@#1:#2::\@nil}
\def\mn@eprint@arXiv#1{\href {http://arxiv.org/abs/#1} {{\tt arXiv:#1}}}
\def\mn@eprint@dblp#1{\href {http://dblp.uni-trier.de/rec/bibtex/#1.xml}
  {dblp:#1}}
\def\mn@eprint@#1:#2:#3:#4\@nil{\def\@tempa {#1}\def\@tempb {#2}\def\@tempc
  {#3}\ifx \@tempc \@empty \let \@tempc \@tempb \let \@tempb \@tempa \fi \ifx
  \@tempb \@empty \def\@tempb {arXiv}\fi \@ifundefined
  {mn@eprint@\@tempb}{\@tempb:\@tempc}{\expandafter \expandafter \csname
  mn@eprint@\@tempb\endcsname \expandafter{\@tempc}}}

\bibitem[\protect\citeauthoryear{{Baran}}{{Baran}}{2012}]{baran2012a}
{Baran} A.~S.,  2012, \actaa, \href
  {http://adsabs.harvard.edu/abs/2012AcA....62..179B} {62, 179}

\bibitem[\protect\citeauthoryear{{Baran}}{{Baran}}{2013}]{baran2013}
{Baran} A.~S.,  2013, \actaa, \href
  {http://adsabs.harvard.edu/abs/2013AcA....63..203B} {63, 203}

\bibitem[\protect\citeauthoryear{{Baran} \& {Winans}}{{Baran} \&
  {Winans}}{2012}]{baran2012b}
{Baran} A.~S.,  {Winans} A.,  2012, \actaa, \href
  {http://adsabs.harvard.edu/abs/2012AcA....62..343B} {62, 343}

\bibitem[\protect\citeauthoryear{{Baran}, {Pigulski}, {Kozie{\l}}, {Og{\l}oza},
  {Silvotti}  \& {Zo{\l}a}}{{Baran} et~al.}{2005}]{baran2005}
{Baran} A.,  {Pigulski} A.,  {Kozie{\l}} D.,  {Og{\l}oza} W.,  {Silvotti} R.,
  {Zo{\l}a} S.,  2005, \mn@doi [\mnras] {10.1111/j.1365-2966.2005.09066.x},
  \href {http://adsabs.harvard.edu/abs/2005MNRAS.360..737B} {360, 737}

\bibitem[\protect\citeauthoryear{{Baran} et~al.,}{{Baran}
  et~al.}{2011}]{baran2011}
{Baran} A.~S.,  et~al., 2011, \mn@doi [\mnras]
  {10.1111/j.1365-2966.2011.18486.x}, \href
  {http://adsabs.harvard.edu/abs/2011MNRAS.414.2871B} {414, 2871}

\bibitem[\protect\citeauthoryear{{Baran}, {Koen}  \& {Pokrzywka}}{{Baran}
  et~al.}{2015a}]{baran2015a}
{Baran} A.~S.,  {Koen} C.,   {Pokrzywka} B.,  2015a, \mn@doi [\mnras]
  {10.1093/mnrasl/slu194}, \href
  {http://adsabs.harvard.edu/abs/2015MNRAS.448L..16B} {448, L16}

\bibitem[\protect\citeauthoryear{{Baran}, {Telting}, {N{\'e}meth}, {Bachulski}
  \& {Krzesi{\'n}ski}}{{Baran} et~al.}{2015b}]{baran2015b}
{Baran} A.~S.,  {Telting} J.~H.,  {N{\'e}meth} P.,  {Bachulski} S.,
  {Krzesi{\'n}ski} J.,  2015b, \mn@doi [\aap] {10.1051/0004-6361/201424877},
  \href {http://adsabs.harvard.edu/abs/2015A%26A...573A..52B} {573, A52}

\bibitem[\protect\citeauthoryear{{Baran}, {Reed}, {{\O}stensen}, {Telting}  \&
  {Jeffery}}{{Baran} et~al.}{2017}]{baran2017}
{Baran} A.~S.,  {Reed} M.~D.,  {{\O}stensen} R.~H.,  {Telting} J.~H.,
  {Jeffery} C.~S.,  2017, \mn@doi [\aap] {10.1051/0004-6361/201629651}, \href
  {http://adsabs.harvard.edu/abs/2017A%26A...597A..95B} {597, A95}

\bibitem[\protect\citeauthoryear{{Charpinet}, {Fontaine}, {Brassard}  \&
  {Dorman}}{{Charpinet} et~al.}{1996}]{charpinet1996}
{Charpinet} S.,  {Fontaine} G.,  {Brassard} P.,   {Dorman} B.,  1996, \mn@doi
  [\apjl] {10.1086/310335}, \href
  {http://adsabs.harvard.edu/abs/1996ApJ...471L.103C} {471, L103}

\bibitem[\protect\citeauthoryear{{Charpinet}, {Fontaine}, {Brassard}, {Chayer},
  {Rogers}, {Iglesias}  \& {Dorman}}{{Charpinet} et~al.}{1997}]{charpinet1997}
{Charpinet} S.,  {Fontaine} G.,  {Brassard} P.,  {Chayer} P.,  {Rogers} F.~J.,
  {Iglesias} C.~A.,   {Dorman} B.,  1997, \mn@doi [\apjl] {10.1086/310741},
  \href {http://adsabs.harvard.edu/abs/1997ApJ...483L.123C} {483, L123}

\bibitem[\protect\citeauthoryear{{Charpinet}, {Fontaine}, {Brassard}  \&
  {Dorman}}{{Charpinet} et~al.}{2000}]{charpinet2000}
{Charpinet} S.,  {Fontaine} G.,  {Brassard} P.,   {Dorman} B.,  2000, \mn@doi
  [\apjs] {10.1086/317359}, \href
  {http://adsabs.harvard.edu/abs/2000ApJS..131..223C} {131, 223}

\bibitem[\protect\citeauthoryear{{Charpinet} et~al.,}{{Charpinet}
  et~al.}{2011}]{charpinet2011}
{Charpinet} S.,  et~al., 2011, \mn@doi [\aap] {10.1051/0004-6361/201016412},
  \href {http://adsabs.harvard.edu/abs/2011A%26A...530A...3C} {530, A3}

\bibitem[\protect\citeauthoryear{{Charpinet}, {Brassard}, {Van Grootel}  \&
  {Fontaine}}{{Charpinet} et~al.}{2014}]{charpinet2014}
{Charpinet} S.,  {Brassard} P.,  {Van Grootel} V.,   {Fontaine} G.,  2014, in
  {van Grootel} V.,  {Green} E.,  {Fontaine} G.,   {Charpinet} S.,  eds,
  Astronomical Society of the Pacific Conference Series Vol. 481, 6th Meeting
  on Hot Subdwarf Stars and Related Objects. p.~179

\bibitem[\protect\citeauthoryear{{Constantino}, {Campbell},
  {Christensen-Dalsgaard}, {Lattanzio}  \& {Stello}}{{Constantino}
  et~al.}{2015}]{constantino2015}
{Constantino} T.,  {Campbell} S.~W.,  {Christensen-Dalsgaard} J.,  {Lattanzio}
  J.~C.,   {Stello} D.,  2015, \mn@doi [\mnras] {10.1093/mnras/stv1264}, \href
  {http://cdsads.u-strasbg.fr/abs/2015MNRAS.452..123C} {452, 123}

\bibitem[\protect\citeauthoryear{{Fontaine}, {Brassard}, {Charpinet}, {Green},
  {Chayer}, {Bill{\`e}res}  \& {Randall}}{{Fontaine}
  et~al.}{2003}]{fontaine2003}
{Fontaine} G.,  {Brassard} P.,  {Charpinet} S.,  {Green} E.~M.,  {Chayer} P.,
  {Bill{\`e}res} M.,   {Randall} S.~K.,  2003, \mn@doi [\apj] {10.1086/378270},
  \href {http://adsabs.harvard.edu/abs/2003ApJ...597..518F} {597, 518}

\bibitem[\protect\citeauthoryear{{Foster}, {Reed}, {Telting}, {{\O}stensen}  \&
  {Baran}}{{Foster} et~al.}{2015}]{foster2015}
{Foster} H.~M.,  {Reed} M.~D.,  {Telting} J.~H.,  {{\O}stensen} R.~H.,
  {Baran} A.~S.,  2015, \mn@doi [\apj] {10.1088/0004-637X/805/2/94}, \href
  {http://adsabs.harvard.edu/abs/2015ApJ...805...94F} {805, 94}

\bibitem[\protect\citeauthoryear{{Ghasemi}, {Moravveji}, {Aerts}, {Safari}  \&
  {Vu{\v c}kovi{\'c}}}{{Ghasemi} et~al.}{2017}]{ghasemi2017}
{Ghasemi} H.,  {Moravveji} E.,  {Aerts} C.,  {Safari} H.,   {Vu{\v c}kovi{\'c}}
  M.,  2017, \mn@doi [\mnras] {10.1093/mnras/stw2839}, \href
  {http://cdsads.u-strasbg.fr/abs/2017MNRAS.465.1518G} {465, 1518}

\bibitem[\protect\citeauthoryear{{Han}, {Podsiadlowski}, {Maxted}, {Marsh}  \&
  {Ivanova}}{{Han} et~al.}{2002}]{han2002}
{Han} Z.,  {Podsiadlowski} P.,  {Maxted} P.~F.~L.,  {Marsh} T.~R.,   {Ivanova}
  N.,  2002, \mn@doi [\mnras] {10.1046/j.1365-8711.2002.05752.x}, \href
  {http://adsabs.harvard.edu/abs/2002MNRAS.336..449H} {336, 449}

\bibitem[\protect\citeauthoryear{{Heber}}{{Heber}}{2016}]{heber2016}
{Heber} U.,  2016, \mn@doi [\pasp] {10.1088/1538-3873/128/966/082001}, \href
  {http://adsabs.harvard.edu/abs/2016PASP..128h2001H} {128, 082001}

\bibitem[\protect\citeauthoryear{{Kern}, {Reed}, {Baran}, {{\O}stensen}  \&
  {Telting}}{{Kern} et~al.}{2017}]{kern2017}
{Kern} J.~W.,  {Reed} M.~D.,  {Baran} A.~S.,  {{\O}stensen} R.~H.,   {Telting}
  J.~H.,  2017, \mn@doi [\mnras] {10.1093/mnras/stw2794}, \href
  {http://cdsads.u-strasbg.fr/abs/2017MNRAS.465.1057K} {465, 1057}

\bibitem[\protect\citeauthoryear{{Kilkenny}, {Koen}, {O'Donoghue}  \&
  {Stobie}}{{Kilkenny} et~al.}{1997}]{Kilkenny1997}
{Kilkenny} D.,  {Koen} C.,  {O'Donoghue} D.,   {Stobie} R.~S.,  1997, \mn@doi
  [\mnras] {10.1093/mnras/285.3.640}, \href
  {http://adsabs.harvard.edu/abs/1997MNRAS.285..640K} {285, 640}

\bibitem[\protect\citeauthoryear{{Koen}, {O'Donoghue}, {Pollacco}  \&
  {Charpinet}}{{Koen} et~al.}{1999}]{koen1999}
{Koen} C.,  {O'Donoghue} D.,  {Pollacco} D.~L.,   {Charpinet} S.,  1999,
  \mn@doi [\mnras] {10.1046/j.1365-8711.1999.02374.x}, \href
  {http://adsabs.harvard.edu/abs/1999MNRAS.305...28K} {305, 28}

\bibitem[\protect\citeauthoryear{{{\O}stensen} et~al.,}{{{\O}stensen}
  et~al.}{2010}]{ostensen2010}
{{\O}stensen} R.~H.,  et~al., 2010, \mn@doi [\mnras]
  {10.1111/j.1365-2966.2010.17366.x}, \href
  {http://cdsads.u-strasbg.fr/abs/2010MNRAS.409.1470O} {409, 1470}

\bibitem[\protect\citeauthoryear{{{\O}stensen} et~al.,}{{{\O}stensen}
  et~al.}{2011}]{ostensen2011}
{{\O}stensen} R.~H.,  et~al., 2011, \mn@doi [\mnras]
  {10.1111/j.1365-2966.2011.18405.x}, \href
  {http://cdsads.u-strasbg.fr/abs/2011MNRAS.414.2860O} {414, 2860}

\bibitem[\protect\citeauthoryear{{{\O}stensen}, {Telting}, {Reed}, {Baran},
  {Nemeth}  \& {Kiaeerad}}{{{\O}stensen} et~al.}{2014}]{ostensen2014}
{{\O}stensen} R.~H.,  {Telting} J.~H.,  {Reed} M.~D.,  {Baran} A.~S.,  {Nemeth}
  P.,   {Kiaeerad} F.,  2014, \mn@doi [\aap] {10.1051/0004-6361/201423611},
  \href {http://adsabs.harvard.edu/abs/2014A%26A...569A..15O} {569, A15}

\bibitem[\protect\citeauthoryear{{Reed} et~al.,}{{Reed}
  et~al.}{2011}]{reed2011}
{Reed} M.~D.,  et~al., 2011, \mn@doi [\mnras]
  {10.1111/j.1365-2966.2011.18532.x}, \href
  {http://adsabs.harvard.edu/abs/2011MNRAS.414.2885R} {414, 2885}

\bibitem[\protect\citeauthoryear{{Reed}, {Foster}, {Telting}, {{\O}stensen},
  {Farris}, {Oreiro}  \& {Baran}}{{Reed} et~al.}{2014}]{reed2014}
{Reed} M.~D.,  {Foster} H.,  {Telting} J.~H.,  {{\O}stensen} R.~H.,  {Farris}
  L.~H.,  {Oreiro} R.,   {Baran} A.~S.,  2014, \mn@doi [\mnras]
  {10.1093/mnras/stu412}, \href
  {http://adsabs.harvard.edu/abs/2014MNRAS.440.3809R} {440, 3809}

\bibitem[\protect\citeauthoryear{{Schindler}, {Green}  \& {Arnett}}{{Schindler}
  et~al.}{2015}]{schindler2015}
{Schindler} J.-T.,  {Green} E.~M.,   {Arnett} W.~D.,  2015, \mn@doi [\apj]
  {10.1088/0004-637X/806/2/178}, \href
  {http://cdsads.u-strasbg.fr/abs/2015ApJ...806..178S} {806, 178}

\bibitem[\protect\citeauthoryear{{Schuh}, {Huber}, {Dreizler}, {Heber},
  {O'Toole}, {Green}  \& {Fontaine}}{{Schuh} et~al.}{2006}]{schuh2006}
{Schuh} S.,  {Huber} J.,  {Dreizler} S.,  {Heber} U.,  {O'Toole} S.~J.,
  {Green} E.~M.,   {Fontaine} G.,  2006, \mn@doi [\aap]
  {10.1051/0004-6361:200500210}, \href
  {http://adsabs.harvard.edu/abs/2006A%26A...445L..31S} {445, L31}

\bibitem[\protect\citeauthoryear{{Silvotti} et~al.,}{{Silvotti}
  et~al.}{2014}]{silvotti2014}
{Silvotti} R.,  et~al., 2014, \mn@doi [\aap] {10.1051/0004-6361/201424509},
  \href {http://adsabs.harvard.edu/abs/2014A%26A...570A.130S} {570, A130}

\bibitem[\protect\citeauthoryear{{Telting} et~al.,}{{Telting}
  et~al.}{2012}]{telting2012}
{Telting} J.~H.,  et~al., 2012, \mn@doi [\aap] {10.1051/0004-6361/201219458},
  \href {http://adsabs.harvard.edu/abs/2012A%26A...544A...1T} {544, A1}

\bibitem[\protect\citeauthoryear{{Telting} et~al.,}{{Telting}
  et~al.}{2014}]{telting2014}
{Telting} J.~H.,  et~al., 2014, \mn@doi [\aap] {10.1051/0004-6361/201424169},
  \href {http://cdsads.u-strasbg.fr/abs/2014A%26A...570A.129T} {570, A129}

\bibitem[\protect\citeauthoryear{{Van Grootel}, {Charpinet}, {Brassard},
  {Fontaine}  \& {Green}}{{Van Grootel} et~al.}{2013}]{vangrootel2013}
{Van Grootel} V.,  {Charpinet} S.,  {Brassard} P.,  {Fontaine} G.,   {Green}
  E.~M.,  2013, \mn@doi [\aap] {10.1051/0004-6361/201220896}, \href
  {http://adsabs.harvard.edu/abs/2013A%26A...553A..97V} {553, A97}

\bibitem[\protect\citeauthoryear{{Yong}, {Demarque}  \& {Yi}}{{Yong}
  et~al.}{2000}]{yong2000}
{Yong} H.,  {Demarque} P.,   {Yi} S.,  2000, \mn@doi [\apj] {10.1086/309285},
  \href {http://adsabs.harvard.edu/abs/2000ApJ...539..928Y} {539, 928}

\makeatother
\end{thebibliography}

\bsp	
\label{lastpage}
\end{document}